\documentclass[aps,amsfonts,amsmath,prd,preprint,nofootinbib,tightenlines]{revtex4}

\newcommand{\beq}{\begin{equation}}
\newcommand{\eeq}{\end{equation}}
\newcommand{\erfc}{\mathop{\rm erfc}}
\usepackage{epsfig}

\begin{document}

\title{Cosmic string formation by flux trapping}
 
\author{Jose J. Blanco-Pillado}
\author{Ken D. Olum}
\author{Alexander Vilenkin}

\affiliation{Institute of Cosmology, Department of Physics and Astronomy,\\ 
Tufts University, Medford, MA 02155, USA
}

\def\changenote#1{\footnote{\bf #1}}

\begin{abstract}
We study the formation of cosmic strings by confining a stochastic 
magnetic field into flux tubes in a numerical simulation.  We use 
overdamped evolution in a potential that is minimized when the 
flux through each face in the simulation lattice is a multiple of 
the fundamental flux quantum.  When the typical number of flux 
quanta through a correlation-length-sized region is initially 
about 1, we find a string network similar to that generated by 
the Kibble-Zurek mechanism.  With larger initial flux, the loop 
distribution and the Brownian shape of the infinite strings remain 
unchanged, but the fraction of length in infinite strings is 
increased. A 2D slice of the network exhibits bundles of strings 
pointing in the same direction, as in earlier 2D simulations. 
We find, however, that strings belonging to the same bundle do not 
stay together in 3D for much longer than the correlation length. 
As the initial flux per correlation length is decreased, there is 
a point at which infinite strings disappear, as in the Hagedorn 
transition.
\end{abstract}

\maketitle

\section{Introduction}

Formation of linear defects (strings) at a symmetry-breaking phase
transition is of great interest both in cosmology and in condensed
matter physics.  For a global symmetry breaking, defects are formed
because the phases of the Higgs field (order parameter) are
uncorrelated on scales greater than the characteristic correlation
length $\xi_H$. In the simplest case of a global $U(1)$ symmetry, a
string is formed whenever the phase changes by $2\pi$ around a closed
loop. This is the familiar Kibble-Zurek mechanism
\cite{Kibble,Zurek}. The statistical properties of the resulting
string networks have been studied in numerical simulations
\cite{VV84}, with the following conclusions.

The string network consists of two components: infinite strings and
closed loops. The infinite strings constitute about 80\% of the total
string length and have the shape of random walks; their fractal
dimension is $D=2$, within statistical errors. The number of
closed loops of length between $l$ and $l+dl$ per unit volume is given
by
\beq
n(l)dl = A\xi_H^{-3/2}l^{-5/2}dl,
\label{nl}
\eeq
where $A \sim 1$ is a numerical coefficient. Long Brownian loops of
length $l$ have size $R\sim (\xi_H l)^{1/2}$, and Eq.~(\ref{nl}) gives
a scale-invariant size distribution,
\beq
n(R)dR \sim R^{-4}dR.
\label{nR}
\eeq
Later work showed that these properties are very robust. In
particular, simulations of $Z_2$-string formation, which require a
different symmetry breaking scheme, yield very similar results, except
the fraction of length in closed loops is reduced from $\sim 20\%$ to
$\sim 6\%$ \cite{Kibble86,AEVV86}.

In the case of a gauge symmetry breaking, each string carries a
quantum of magnetic gauge flux, and apart from the Kibble-Zurek
mechanism, an additional mechanism of string formation can operate
\cite{HR00}.  Any magnetic gauge field present before the phase
transition will tend to be squeezed into quantized flux tubes after
the phase transition.  This mechanism may operate in superconductors,
where the stochastic magnetic field can be produced by thermal
fluctuations, and in cosmological phase transitions, where the field
can be due either to thermal or to quantum fluctuations. Flux trapping
becomes the dominant mechanism of string formation when the magnetic
field fluctuations get sufficiently large, so that the typical area
over which the magnetic flux is equal to one flux quantum is
smaller than $\xi_H^2$ \cite{HR00}.  This condition
is often satisfied in superconductors and may well be satisfied in a
cosmological setting.

Another interesting example where flux trapping may be important is
the brane inflation model. Inflation in this model is driven by the
attractive interaction energy of a $D3$ and an anti-${D3}$ brane
separated in extra dimensions \cite{DT}.  The branes eventually
collide and annihilate, and cosmic $D$-strings ($D1$-branes) can be
produced in the process \cite{Tye,Tye2,DV04,CMP04}.  It has been
argued in \cite{DV04} that $D$-string formation by Kibble-Zurek
mechanism is strongly suppressed in this model. However, a string
network may still be formed by flux trapping.  Quantum fluctuations of
the gauge fields living on the branes get strongly amplified in the
process of brane annihilation \cite{DV03}.  The resulting magnetic
fields can then be squeezed into quantized flux tubes to become
$D$-strings.  Fundamental ($F$) strings can be formed in a similar
manner, by squeezing the electric component of another gauge field
(orthogonal to that responsible for $D$-strings) into electric flux
tubes.\footnote{If both $F$ and $D$ strings are present, they can form
bound states and combine into an $FD$ network \cite{CMP04,DV04}.}

String networks formed by flux trapping may be rather different from
those formed by the Kibble-Zurek mechanism.  An important parameter
here is the rms magnetic flux $\Phi_c$ through an area $\xi_B^2$, where 
$\xi_B$ is the correlation length of the magnetic field.  If $\Phi_c$
is much greater than one flux quantum, then the strings will form in
bundles containing many strings each \cite{HR00,SBZ02,KR03,DKR04}. This
phenomenon has been observed in 2-dimensional simulations, where
vortices (and anti-vortices) had a tendency to bundle together for
large values of $\Phi_c$.  However, the properties of 3-dimensional
string networks formed by flux trapping remain largely unknown.  What
are the shapes of long strings?  Are they still Brownian?  What
fraction of the total length is in closed loops and what is their size
distribution?  If long strings form bundles, do they tend to stay in
the same bundle, or switch from one bundle to another?  To address
these questions, we have developed a numerical simulation of string
formation by flux trapping.

Simulations of detailed Higgs and gauge field dynamics in a phase
transition are computationally expensive, and the maximum simulation
size that we could achieve in this way would not be sufficient to get
a good handle on the statistical properties of string networks.  We
therefore took a different approach.  Our starting point is a
stochastic magnetic field, which has presumably originated from either
thermal or quantum fluctuations.  We put the field on a lattice and
use a relaxation technique to relax it to a state in which the flux
through each plaquette is an integer multiple of the flux quantum.
Strings can then be easily traced by following the flux lines from one
cell to another.  The net flux into each lattice cell remains
equal to zero throughout the relaxation process, ensuring the
continuity of strings.  A ``practical'' application of these
simulations is that they can be used as initial conditions for
dynamical simulations of evolving string networks.

\section{Simulation}

\subsection{Initial Conditions}

We have performed a series of numerical simulations borrowing
techniques previously used in lattice field theory (see for example
Ref.~\cite{BPO}).  We set up an initial configuration
for the magnetic field on a cubic lattice by assigning the values of
the vector field ${\bf A}({\bf x})$ to the lattice links.  It is then
straightforward to compute the magnetic field flux through an
individual plaquette from the values of the vector field on its
boundaries. Since we are only interested in the magnetic part of the
gauge field, it is sufficient to consider a vector potential of the 
form
\beq
\label{A}
{\bf A}({\bf x}) = \sum_{{\bf k}} 
(2V\omega_{\bf k})^{-1/2}{\bf a}({\bf k})e^{i{\bf k}\cdot{\bf x}},
\eeq
where $\omega_{\bf k} = |{\bf k}|$, $V=L^3$ is the volume of the simulation
box, and we use the system of units in which $\hbar=c=1$. The reality
of the electromagnetic field is ensured by the constraint,
${\bf a}({\bf -k})= {\bf a^*}({\bf k})$.

We impose periodic boundary conditions, so the wave vector ${\bf k}$
takes a discrete set of values, ${\bf k}=(2\pi/L) {\bf
n}=(2\pi/L) (n_1,n_2,n_3)$ with $n_1,n_2,n_3 = 0, \pm 1,...$.
The number of lattice points along each side of the box is $N=L/\Delta
x$, where $\Delta x$ is the lattice spacing. Wavelengths shorter than
$2\Delta x$ cannot be represented in such a lattice, so we cut off the
summation in (\ref{A}) at $n_j =\pm N/2$,

\beq
{\bf A}({\bf x_{\bf m}}) = \sum_{\bf n}
(4 \pi L^2 |{\bf n}|)^{-1/2}{\bf a}({\bf n})e^{2 \pi i {\bf n} \cdot {\bf m}/N}.
\eeq
Here, we have denoted the points on the lattice by ${\bf x_{\bf m}}=
\Delta x\,{\bf m} = \Delta x\,(m_1,m_2,m_3)$.

We obtain particular realizations of the vector field
by drawing the values of the coefficients ${\bf a}({\bf
k})$ from an appropriate Gaussian distribution. In the case
of superconductors, the distribution is expected to have a thermal
(Rayleigh-Jeans) form, with a cutoff at short wavelengths \cite{DKR04},
\beq
\label{aspectrum}
\langle a_i^*({\bf k}) a_j({\bf k'}) \rangle = 
\frac{K}{\omega_{\bf k}} e^{-(k/k_c)^2}
\delta_{ij}\delta_{{\bf kk}'}, 
\eeq
where $k_c$ is related to the dissipation rate of the magnetic field,
and $K$ gives the amplitude of the spectrum, which in the thermal case
is just the temperature.
Defining the cutoff wavelength $\lambda_c = 2 \pi/k_c$, we can
rewrite this as
\beq
\langle  a_i^*({\bf n}) a_j ({\bf n'}) \rangle =
\frac{K L}{2\pi|{\bf n}|}e^{-\left(\lambda_c |\bf{n}|/L\right)^2}
\delta_{ij}\delta_{{\bf nn}'}.
\label{spectrum}
\eeq

This form of the spectrum for gauge field fluctuations has also been
found in brane inflation models, with $k_c \sim K \sim
1/\tau$, where $\tau$ is the characteristic timescale of brane
annihilation \cite{DV03}.

\subsection{Relaxation technique}

Once the initial conditions for the magnetic field have been set, we
need a mechanism that confines the field into
strings carrying a fundamental unit of flux,
\beq
\Phi_0 = e^{-1} \equiv \alpha^{-1/2},
\eeq
where $e$ is the gauge coupling.  Our final state should then have the
magnetic flux through all plaquettes equal either to zero or to an
integral multiple of $\Phi_0$.  This will allow us to follow the flux
around the simulation box to find the corresponding network of
strings.

In Appendix \ref{sec:N}, we compute the the rms magnetic flux $\Phi_c$ through
a disk of diameter $\lambda_c$ in the spectrum of
Eq.~(\ref{spectrum}).  We define the rms number of flux quanta in such a disk,
\beq
{\cal N}= \Phi_c/\Phi_0 \approx 0.46(K\alpha\lambda_c)^{1/2}\,.
\label{N}
\eeq 
For ${\cal N}>1$, strings are expected to form bundles with
several strings per bundle.  The typical number of strings in a bundle
is larger than ${\cal N}$, because bundles appear not in random places
but where the flux in a given direction is larger than usual.

As mentioned before, the vector field ${\bf A}$ lives naturally on
the links of the lattice and enters the computation of the magnetic
flux through the adjacent plaquettes.  It is clear that changing the
value of the gauge field on a link separating two plaquettes will only
displace part of the flux from one of those plaquettes to the other,
keeping the total flux unchanged. We would like to implement an
algorithm that will drive the magnetic flux through each plaquette to
one of the values $\Phi=n\Phi_0$ with $n = 0, \pm 1, ...$.  We
accomplish this by evolving the vector field $A_l$ on each link $l$
according to the following equation of motion,
\beq
\label{relax-method}
\frac{dA_l}{dt} = - \sum_{l \in f} \frac{\partial V(\Phi_f)}{\partial A_l}\,,
\eeq
where the sum is performed over all four plaquettes that have $l$ as
a side and $V(\Phi_f)$ is a function of the magnetic flux
$\Phi_f$ through the $f$-th plaquette that has its minima at $\Phi =
n\Phi_0$.  (It is assumed that $V(-\Phi)=V(\Phi)$, so $V(\Phi)$
is independent of how one chooses the direction normal to the
plaquette.)  Eq.~(\ref{relax-method}) describes an overdamped
system which is driven to the minimum of the potential energy,
\beq\label{potential-energy}
U = \sum_f V(\Phi_f),
\eeq
where the summation is taken over all the plaquettes in the lattice.  The form of the
function $V(\Phi)$ that we used in the simulation is given in Appendix
 \ref{sec:potential}. This particular form is not essential for the method to work,
although it is important to make sure that the final flux through a
plaquette is approached in a smooth and monotonic way. This is
achieved by choosing the time step $\Delta t$ of the
simulation sufficiently small.

We finally have to choose a suitable lattice spacing $\Delta x$. A
natural choice is to set it equal to the correlation length, $\Delta x
= \lambda_c$, as it was done in earlier simulations of string
formation by the Kibble mechanism \cite{VV84,Kibble86,AEVV86}.  This works
fine when the parameter ${\cal N}=\Phi_c/ \Phi_0 \lesssim 1$, but for
large values of ${\cal N}$ there is a large number of strings per
lattice cell, which makes it hard to resolve individual strings.
Hence, we used
\beq
\Delta x \sim \begin{cases}\lambda_c & {\cal N} \le 1 \\
\lambda_c\,  {\cal N}^{-1/2} & {\cal N} > 1\,.
\end{cases}
\label{DELTAX}
\eeq

The overdamped evolution of Eq.\ (\ref{relax-method}) converges on a
state where the ``potential energy'' of Eq.\ (\ref{potential-energy})
vanishes because the flux is properly quantized.  However, this
evolution may be quite slow, because some plaquettes may have flux
nearly equidistant between two multiples of $\Phi_0$ and so have
little force driving them toward one or another.

At a phase transition that confines flux into strings, a loop of flux
can also shrink down to nothing.  (For example, if there is a loop
with less than half a quantum of flux, this must happen.)  In general,
in a region with no net flux to the outside (as we have because of
periodic boundary conditions), in a slow phase transition the flux
will tend to diffuse to nothing, whereas in a rapid transition it will
tend to be quantized.  In the simulation, this choice is affected by
the size of $\Delta x$.  If $\Delta x$ is taken so small that the
average number of flux quanta through each plaquette is much less than
1, then the algorithm takes a long time to assemble a whole quantum of
flux, and the flux tends to dissipate, so that there are very few
strings.  Thus even if ${\cal N}$ is large, if we make $\Delta x$ small we get
primarily dissipation.  Since we are interested in studying the shape
of the string network, we will not make this choice.  However, when
${\cal N}$ is small, even if we choose $\Delta x \sim \lambda_c$ as in
Eq.~(\ref{DELTAX}), the average flux per plaquette will still be
small, and so the diffusion of the flux may become important.

\section{Results}

\subsection{Large ${\cal N}$}

\begin{figure}
\epsfig{file=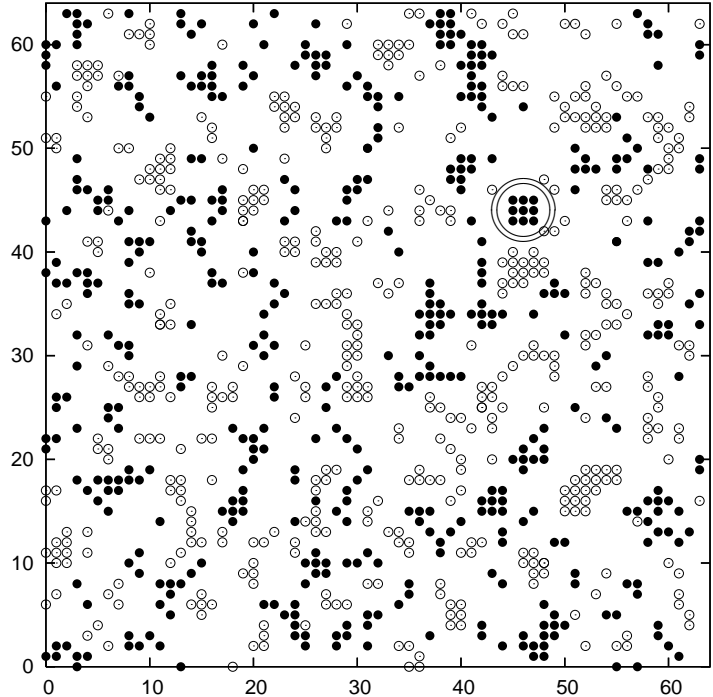,width=10cm,angle=-90}
\put(-88,-87.5){\circle{20}}
\put(-88,-87.5){\circle{22}}
\caption{Two-dimensional slice of a $64^3$ lattice simulation. The black
and white circles represent the position of vortices and anti-vortices respectively. 
Clusters of vortices (anti-vortices) are clearly visible in this figure, indicating 
that the corresponding strings in the 3D lattice form bundles that pierce this surface 
with the same orientation. We encircled the particular group of vortices that we study
in detail in Fig.~\ref{STRING-BUNCH}.}
\label{BUNCHES}
\end{figure}

We are mainly interested in studying the properties of a string network
with a reasonably large number of strings per bundle.
We have generated initial conditions for this type of configuration by
setting $\lambda_c = 6 \Delta x$ and ${\cal N} \sim 3.2$.  The
result of the relaxation procedure shows that a typical 2D slice of
the final configuration has several bunches, some of them
including 10 strings or more.  We show in Fig.~\ref{BUNCHES}
a 2D slice of a $64^3$ lattice simulation As one might expect, it
resembles the 2D simulations discussed in Refs.~\cite{SBZ02,KR03}.

\begin{figure}
\epsfig{file=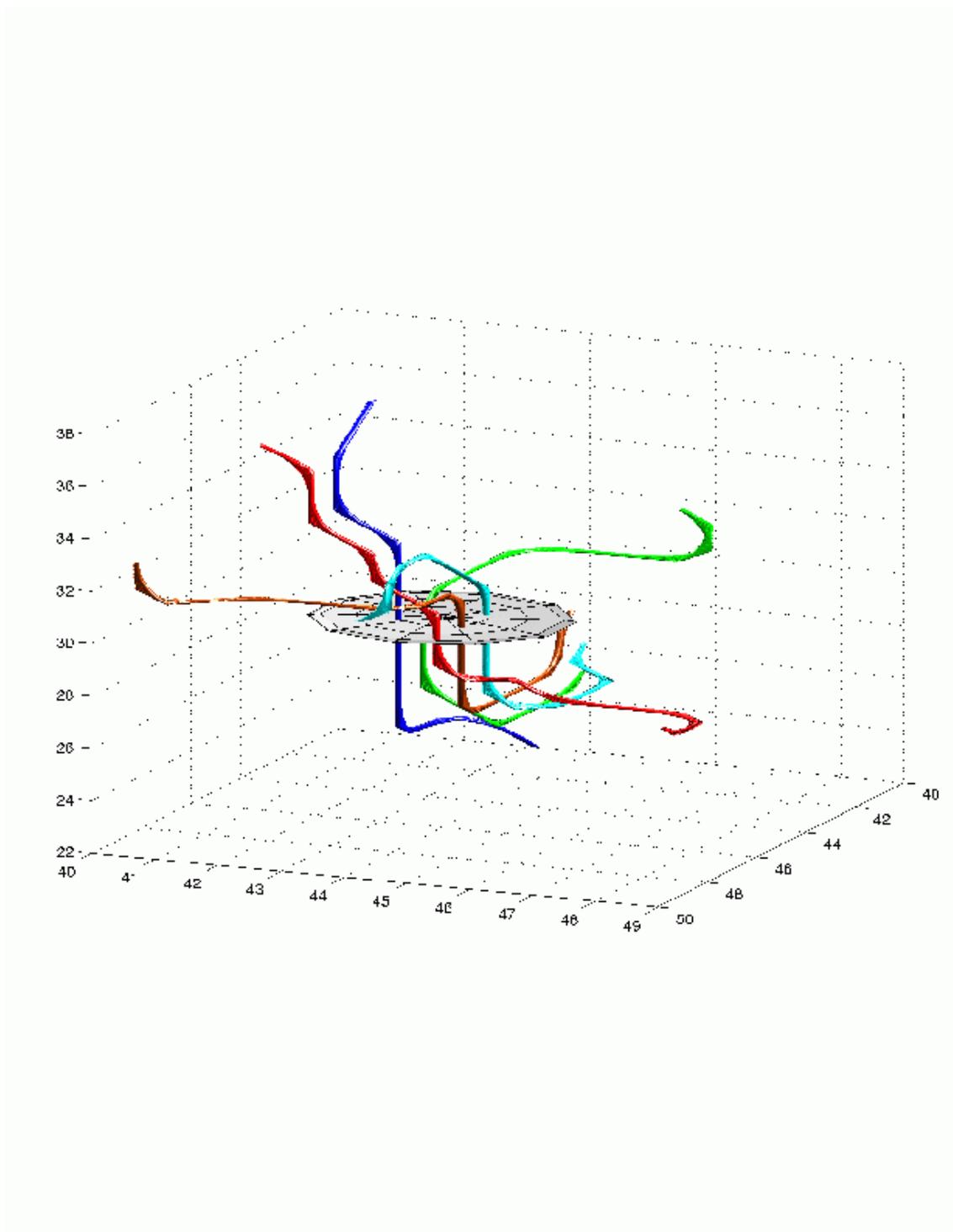,width=15cm}
\caption{3D structure of 5 of the 9 strings encircled in Fig.~\ref{BUNCHES}.
We also plot a 
small portion of the 2D slice to indicate the region of space where
the strings come together as they go through that plane.}
\label{STRING-BUNCH}
\end{figure}

To find out how long the string bunches stay together, we followed the
three-dimensional trajectories of strings for a few steps in both
directions, starting from the bunch encircled in Fig.~\ref{BUNCHES}.
The result is plotted in Fig.~\ref{STRING-BUNCH}.
We see that the strings only stay
together for a few steps, quickly branching off in different
directions (and possibly joining other bunches).

\begin{figure}
\epsfig{file=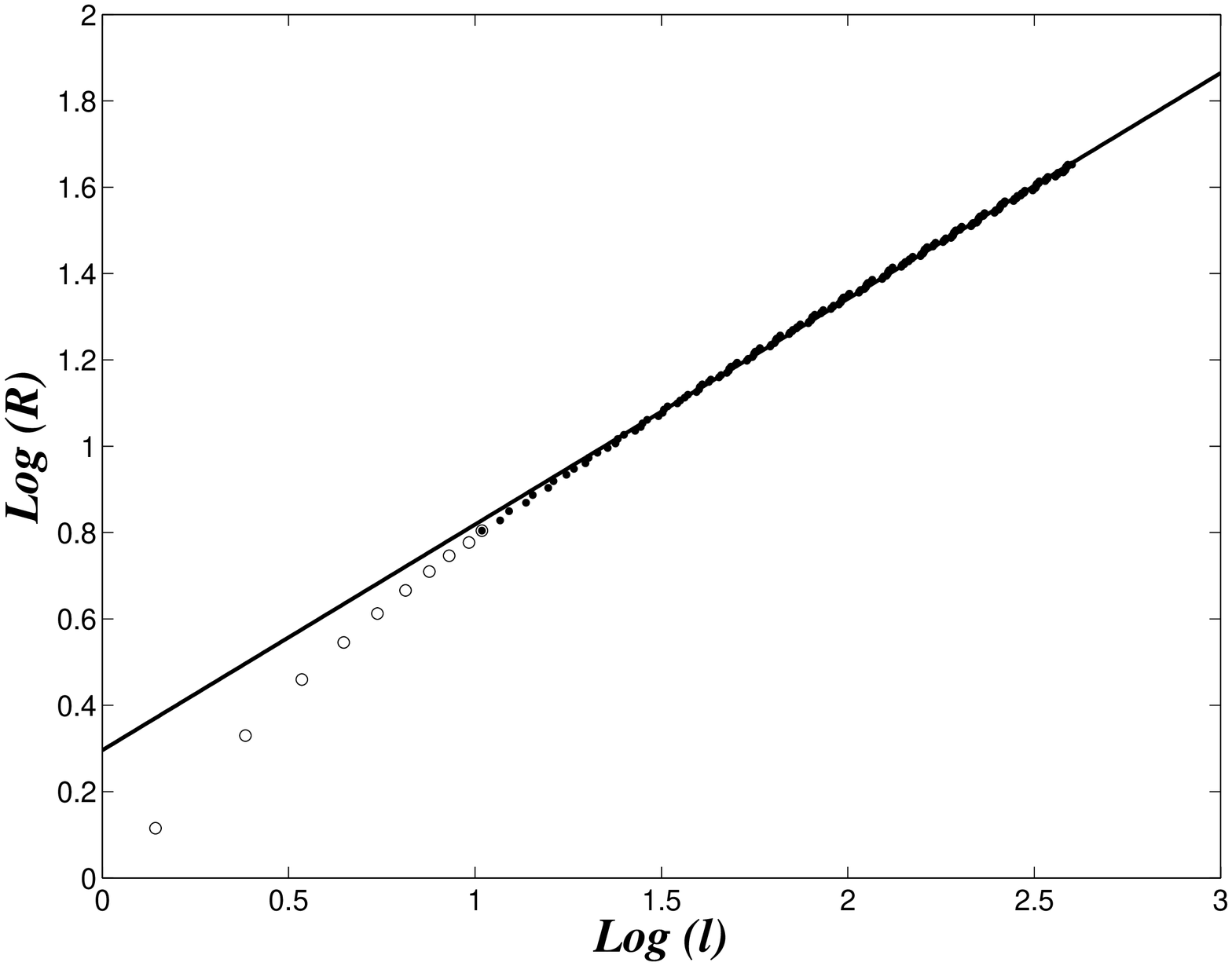,width=13cm}
\caption{Average displacement between two points on the 
string versus the corresponding length along the string for the case 
with ${\cal N}\sim 3.2$.  Here and below white circles indicate points which
are not included in the numerical fit.}
\label{RL-LARGE}
\end{figure}

We have also computed the fractal dimension of strings by averaging
the distance $R$ between points separated by length $l$ along the
string. We have computed this function $R(l)$ by averaging over all
sufficiently long strings present in $30$ realizations; the
result is shown in Fig.~\ref{RL-LARGE}.

The best fit to the data is given by
\beq
R = Al^{\beta},
\label{R(l)}
\eeq
with
\beq 
A = 2.00 \pm 0.05 \qquad \beta = 0.52\pm 0.01,
\eeq
where all lengths are measured in units of $\Delta x$.  The stated
errors here and below are an indication of the uncertainty in the
parameter values, based on differences between separate
runs.  Equation (\ref{R(l)}) is consistent with a random walk shape,
$\beta= 1/2$, of the strings at large $R$.  The
effective step of the random walk, $\xi\sim A^2 \sim 4$, is comparable
to $\lambda_c$, as expected.  It is also comparable to the
characteristic length for which strings in a bundle stay together.

Since our simulations were performed in a box with periodic boundary
conditions, all the strings found in the lattice are, in fact, closed
loops.  The infinite string component of the network is represented by
very long strings, which wind many times around the box.  We counted
all strings shorter than $2L$ as loops and the rest as infinite
strings, but our results are insensitive to the particular
choice of $2L$ as the separation between these two components of the
network.  We found that the fraction of energy stored in infinite
strings is about $98\%$ versus $2\%$ in closed loops.

\begin{figure}
\epsfig{file=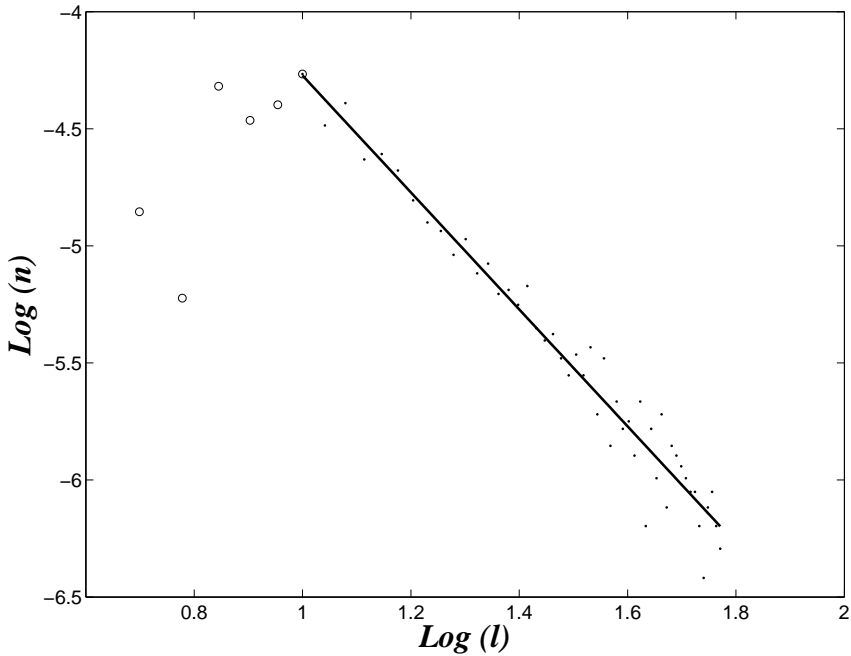,width=13cm}
\caption{Spectrum of closed string loops as a function of their 
length for the case with ${\cal N}\sim 3.2$. }
\label{SPECTRUM-LARGE}
\end{figure}

The length distribution of loops is shown in
Fig.~\ref{SPECTRUM-LARGE}.
The best fit to the data for sufficiently large loop sizes is
\beq
n(l)dl = Bl^{-\gamma}dl,
\label{n(l)}
\eeq
with the following parameters,
\beq 
B = 0.017 \pm 0.001 \qquad \gamma = 2.50\pm 0.03
\eeq
Once again, this is consistent with the scale-invariant distribution
of Eqs.\ (\ref{nl},\ref{nR}).

We note that a string network with somewhat similar properties has
been obtained by Pogosian and Vachaspati from a very different kind 
of simulation \cite{PV97}. They used a random phase distribution on 
a lattice, as in the Vachaspati-Vilenkin (VV) method \cite{VV84}, but 
introduced additional phase differences $2\pi n$ along the links, 
with the integer $n$ taken from a Gaussian distribution. If the rms 
value of $n$ is large, $n_{rms} \gg 1$, then the typical number of 
strings through a plaquette is also large, and one can expect a 2D 
slice of the lattice to exhibit bunches, with about $n_{rms}$ strings 
per plaquette.  The number of closed loops is reduced compared to the 
standard  VV simulation, because when a string returns to the same
cell it started from, it does not necessarily close back on itself.
Instead, it can now be connected to any one of $\sim n_{rms}$ strings 
passing through the cell.

\subsection{${\cal N}\sim 1$}

To simulate a network with ${\cal N}\sim 1$, we have chosen the
following parameters: $\lambda_c = \Delta x$, ${\cal N} \sim 0.28$. 
These values have been selected to adjust the rms magnetic flux
through a plaquette to $0.3\Phi_0$. This roughly corresponds to the
probability of $\sim 30\%$ of having a string through a plaquette in
Kibble mechanism simulations of \cite{VV84}. With this set of
parameters, the magnetic fluxes through neighboring plaquettes are
only weakly correlated with one another, so we expect that strings
will not form bundles and that the resulting network will be similar to
the ones obtained in \cite{VV84}.  And indeed, the results we obtained
are nearly identical to those of \cite{VV84}, even though the
numerical methods used in the two simulations are rather different in
nature.

We have performed these simulations in cubic lattices of size $16^3$,
$32^3$ and $64^3$ and found that the average fraction of string length
in long (infinite) strings was respectively $86\%$, $84\%$ and
$82\%$. These values are very close to those found in
Ref.~\cite{VV84}.

\begin{figure}
\centering\leavevmode
\epsfig{file=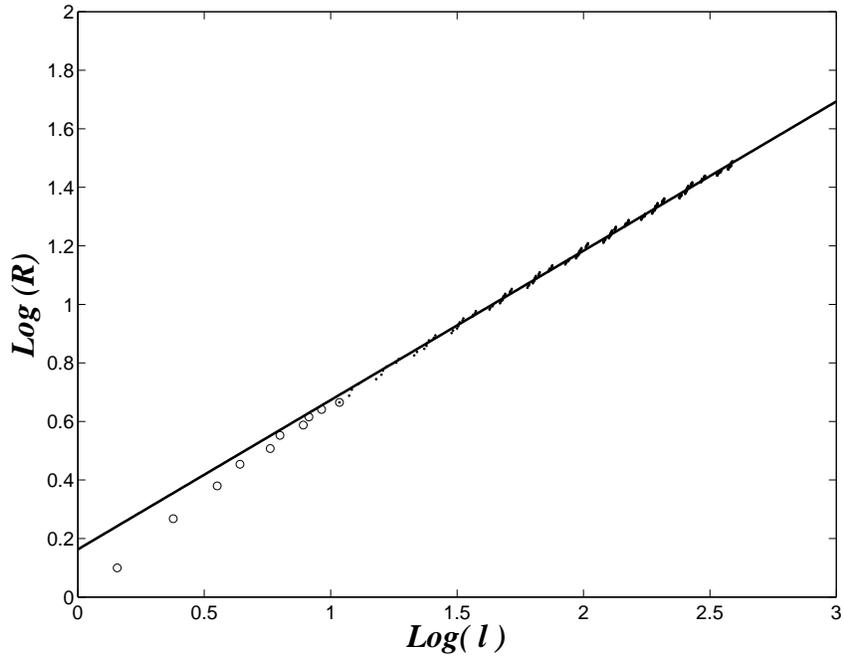,width=13cm}
\caption{
Average displacement between two points on the string versus the 
corresponding length along the string for the case 
with ${\cal N}\sim 0.28$.
}
\label{RL-SMALL}
\end{figure}

\begin{figure}
\centering\leavevmode
\epsfig{file=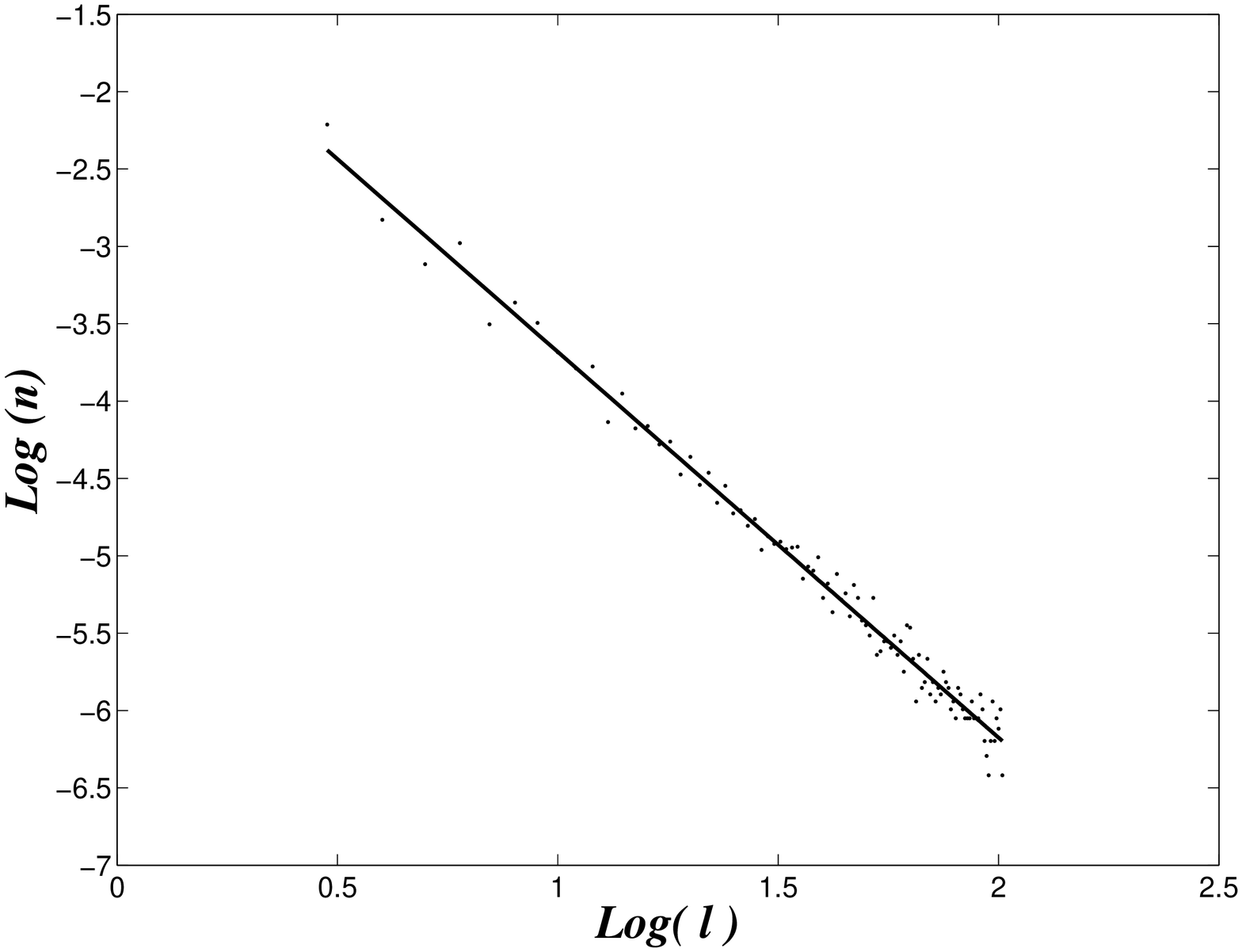,width=13cm}
\caption{Spectrum of closed string loops as a function of their 
length for the case with ${\cal N}\sim 0.28$.}
\label{SPECTRUM-SMALL}
\end{figure}

The distance versus length along the string, $R(l)$, and the length
distribution of loops, $n(l)$, are plotted in Figs.~\ref{RL-SMALL}
and \ref{SPECTRUM-SMALL},
respectively, for that largest box of size
$64^3$. The data are well fitted by Eqs.~(\ref{R(l)}),(\ref{n(l)}) with
\beq 
A= 1.45 \pm 0.03 \qquad \beta = 0.51\pm 0.01,
\eeq
\beq 
B = 0.065 \pm 0.003 \qquad \gamma = 2.51\pm 0.02.
\eeq
As before, the strings have Brownian shapes and the loops exhibit a
scale-invariant distribution.

\subsection{Small ${\cal N}$}

\begin{figure}
\centering\leavevmode
\epsfig{file=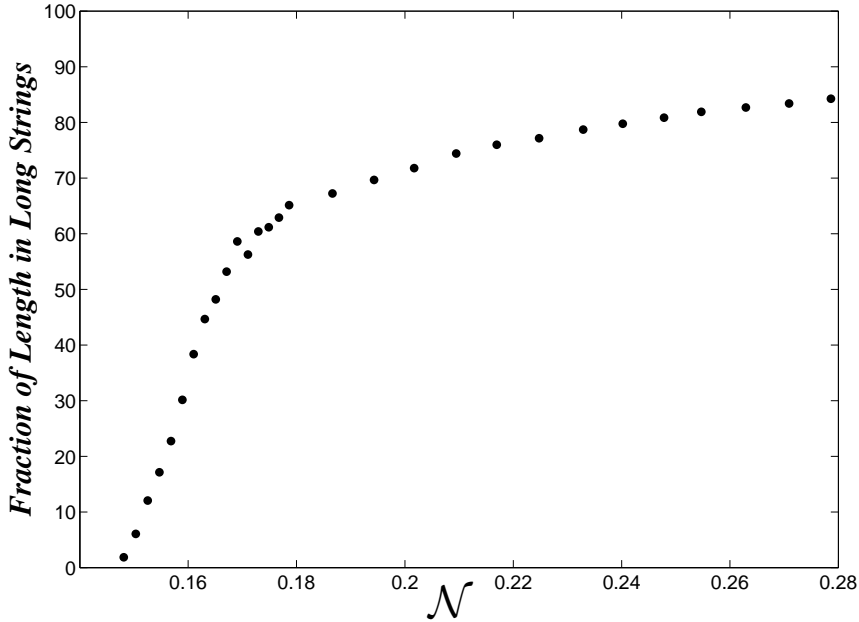,width=13cm}
\put(-190,10){\Large {\bf {$\cal{N}$}}}
\caption{Fraction of the total string length in infinite strings as 
a function of the parameter ${\cal N}$.}
\label{SMALL-N-TRANSITION}
\end{figure}

To explore the properties of the network for ${\cal N}\ll 1$, we kept
$\lambda_c =\Delta x$ and gradually decreased ${\cal N}$,
starting from the value of $0.28$ cited in the preceding
subsection.  For ${\cal N}\ll 1$, the rms flux through a plaquette is
much smaller than $\Phi_0$.  As a result, most plaquettes have no
strings.  The strings have the form of small closed loops and appear
in rare places where the magnetic field fluctuates well above its rms
value.

\begin{figure}
\centering\leavevmode
\epsfig{file=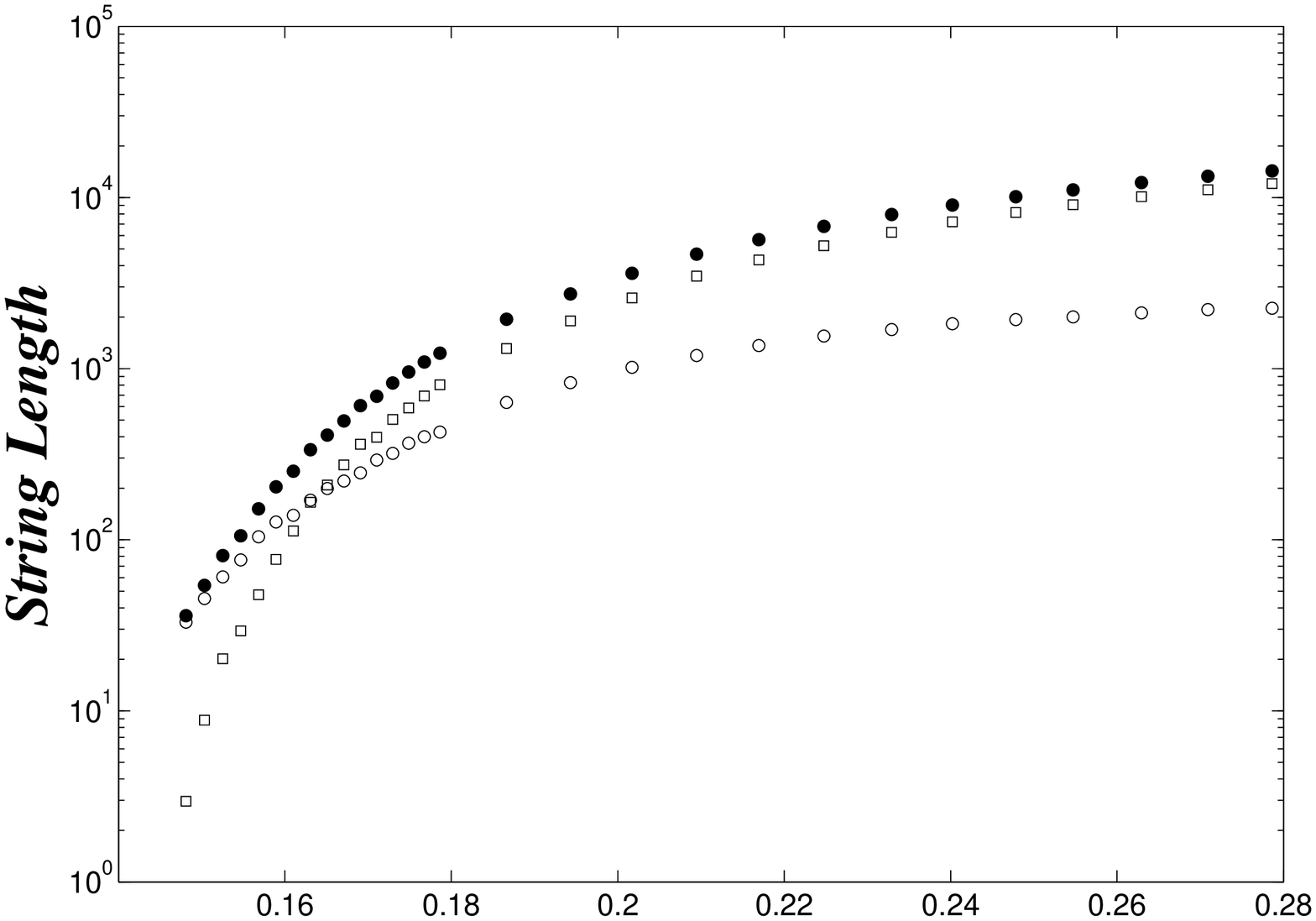,width=13cm}
\put(-190,5){\Large {\bf {$\cal{N}$}}}
\caption{Plot of the average total string length (black circles) as well 
as the different contributions from infinite (squares) and closed
strings (white circles) as a function of ${\cal N}$. The average is 
performed over $100$ simulations of a $32^3$ lattice.}
\label{FIG8}
\end{figure}

In Fig.~\ref{SMALL-N-TRANSITION} we plot the fraction of length in 
infinite strings as a function of ${\cal N}$.  The simulations
discussed in subsections III.A and III.B above correspond to 
${\cal N}= 3.2$ and ${\cal N}=0.28$, respectively.  As ${\cal N}$ 
is decreased, infinite strings constitute
a smaller and smaller fraction of the total string length and finally
disappear at ${\cal N}_c\approx 0.15$. The total string energy density
also decreases with the decrease of ${\cal N}$ (see Fig.~\ref{FIG8}), but
remains finite at (and below) the critical value ${\cal N}_c$.  
For all values of ${\cal N}>{\cal N}_c$ the loop distribution
retains its scale-invariant form (\ref{nl}).  For ${\cal N}<{\cal
N}_c$, most of the string length is in the smallest loops, and the
number density of loops $n(l)$ decreases rapidly with the loop's
length $l$.

It is interesting to note that very similar behavior is observed in a
thermal ensemble of strings with a lower cutoff on the loop length,
$l_{min}$. As the string length per unit volume is decreased below a
certain critical value, $\rho_c\sim l_{min}^{-2}$, the system
undergoes the Hagedorn transition, characterized by the disappearance
of infinite strings. It was noted earlier \cite{MT87,SV88} that string
configurations resulting from simulations of Kibble-Zurek-type phase
transitions resemble those obtained from a thermal string ensemble.

However, we also note that in this regime, the average flux per
plaquette is quite small.  This appears to increase the effects of
diffusion of the flux as compared to confinement, and so it is
possible that the decrease in long strings is due to the increased
diffusion.

\section{Conclusions}

We have studied the statistical properties of string networks formed
by the flux trapping mechanism.  As anticipated in earlier analyses
\cite{HR00,SBZ02,KR03,DKR04}, the character of the network depends on 
the rms flux through a correlation-size region, $\Phi_c$, or
equivalently on the parameter ${\cal N}=\Phi_c/\Phi_0$, where $\Phi_0$
is the flux quantum.

We found that for large values of ${\cal N}$ most of the string length
is in a tangled network of infinite strings, having the shape of
random walks, with closed loops contributing only a small fraction of
the total length.  The step of the random walk is comparable to the
correlation length $\lambda_c$ of the magnetic field.  The typical
inter-string separation, $d\sim \lambda_c/\sqrt{\cal N}$, can be much
smaller.  A 2D slice of the simulation exhibits bundles of strings
pointing in the same direction, with $\sim{\cal N}$ strings per
bundle.  We found, however, that strings belonging to the same bundle
do not stay together for long.  As we follow them in 3D, they part
each other's company on a length scale not much exceeding $\lambda_c$.

For ${\cal N}\sim 1$, there are no string bundles, and the properties
of the string network are nearly identical to those of the networks
formed by the Kibble-Zurek mechanism.

For small values of ${\cal N}$, the string density is exponentially
suppressed and most of the string length is in the form of small loops
of length $l\sim\lambda_c$.

How does the initial string configuration affect the subsequent
evolution of the network?  For ${\cal N}\gg 1$, the properties of the
network differ from the standard initial configuration of
Ref.~\cite{VV84} on relatively short length scales, ${\cal
N}^{-1/2}\lambda_c \lesssim l \lesssim \lambda_c$, which can have an 
effect on early string evolution.  These effects can be studied
by dynamical string simulations in an expanding universe, with initial
conditions generated by simulations of the type described here. At
late times, the string bundles break apart, and we expect the network
evolution to follow the standard scenario (see, e.g., \cite{Book}).
For type-I strings, which can have arbitrary large winding numbers, the
initial configuration with ${\cal N}\gg 1$ may facilitate the formation
of higher-winding strings\footnote{Field theory simulations \cite{DA}
indicate that this is indeed the case, although much
bigger simulations are needed in order to obtain string bunches with 
${\cal N}\gg 1$.}. This may have an effect on the subsequent
evolution of the network. For ${\cal N}\sim 1$, the initial configuration is similar to the
standard one, so the standard evolution scenario is followed from the
very beginning. Finally, for ${\cal N}\ll 1$, all strings are in the form of small
closed loops, which rapidly shrink and disintegrate.

\section*{ACKNOWLEDGMENTS}

This work was supported in part by the National Science Foundation
under grants 0457456 and 0353314.

\appendix 

\section{Average flux through a disk}
\label{sec:N}
We want to compute the rms magnetic flux through a disc of radius
$R$. We follow similar calculations done in \cite{KR03, DKR04}. The mean square flux is
\beq
\langle \Phi(R)^2\rangle = \left\langle \int_0^R{d^2x d^2y B_z(x,z=0) B_z(y,z=0)} \right\rangle ~.
\eeq
Using our definition for ${\bf A}$ given in Eq.~(\ref{A}),
this becomes
\begin{eqnarray*}
\langle \Phi(R)^2 \rangle=
\int_0^R{d^2x d^2y   \sum_{{\bf q},{\bf k}}
(4V^2 \omega_{\bf q} \omega_{\bf k})^{(-1/2)}
\left[q_x  k_x \langle a^*_y({\bf q}) a_y({\bf k}) \rangle 
+  q_y  k_y \langle a^*_x({\bf q}) a_x({\bf k})    \rangle 
\right]
e^{i({\bf k}\cdot{\bf x}-{\bf q}\cdot{\bf y})}}\,.
\end{eqnarray*}
By ${\bf x}$ and ${\bf y}$ we mean the 3-vectors whose $z$
component is 0, and we have neglected the cross terms that average to
zero. Using now equation (\ref{aspectrum}) we arrive at
\beq
\langle \Phi(R)^2 \rangle=
\int_0^R{d^2x d^2y   \sum_{{\bf k}}
(2 V \omega_{\bf k})^{-1}
(k_x^2 +k_y^2)\frac{K}{\omega_{\bf k}}
e^{-(k/k_c)^2}e^{i{\bf k}\cdot({\bf x}-{\bf y})}}\,,
\eeq
where $k$ is the length of ${\bf k}$.
We now take the infinite volume limit and replace the sum by an
integral, using
\beq
\frac{1}{V} \sum_{{\bf k}} \rightarrow \frac{1}{{(2 \pi)}^3} \int d^3 k\,,
\eeq
so we can rewrite the expression above in the following form,
\beq
\langle \Phi(R)^2 \rangle= \frac{K}{2(2\pi)^3}
\int_0^R{ d^2x d^2y \int{ d^3k\frac{k_\perp^2}{k^2} 
e^{-(k/k_c)^2}e^{i{\bf k}\cdot({\bf x}-{\bf y})}}}\,,
\eeq
where we have defined $k_\perp=(k_x^2+ k_y^2)^{1/2}$. We can exchange the
order of the integrals to get,

\begin{eqnarray*}
\langle \Phi(R)^2 \rangle &=&
\frac{K}{2(2\pi)^3}
\int{d^3k\frac{k_\perp^2}{k^2}e^{-(k/k_c)^2}
\left(\int_0^R{r dr \int_0^{2\pi}{d\theta 
e^{i k_\perp r \cos\theta}}}\right)^2}\\ 
&=& \frac{K}{4\pi}
\int{d^3k\frac{k_\perp^2}{k^2}}e^{-(k/k_c)^2}
\left(\int_0^R{ r dr J_0(k_\perp r)}\right)^2\,.
\end{eqnarray*}
Performing the integral in $r$ we obtain
\beq
\langle \Phi(R)^2 \rangle =
\frac{KR^2}{4\pi}
\int\frac{d^3k}{k^2}e^{-(k/k_c)^2}J_1(k_\perp R)^2
\eeq
We now break up the integral into an integral in $k_z$, a radial
integral in $k_\perp$, and a trivial angular integral, to get
\begin{eqnarray*}
\langle \Phi(R)^2 \rangle &=&
\frac{KR^2}{2}
\int_0^\infty dk_\perp\, k_\perp J_1(k_\perp R)^2
e^{-(k/k_c)^2}\int dk_z \frac{e^{-(k_z/k_c)^2}}{k_z^2 + k_\perp^2} \\ 
&=& \frac{\pi K R^2}{2} \int_0^\infty dk_\perp J_1(k_\perp R)^2\erfc(k/k_c)
= \frac{\pi K R}{2}{\cal F}\left(\frac{\lambda_c}{R}\right)\,,
\end{eqnarray*} 
where
\beq
{\cal F}(z) = \int_0^\infty{dy\,J_1(y)^2\erfc\left(\frac {y z} {2\pi}\right)}
= \frac{2\pi^{5/2}}{3z^3}\,{}_2F_2\left(\frac{3}{2},\frac{3}{2},\frac{5}{2},
3, -\frac{4\pi^2}{z^2}\right)
\eeq
where ${}_2F_2$ is a hypergeometric function.

We can now define ${\cal N}$ as the rms flux through a disk of
diameter $\lambda_c$, i.e., $R = \lambda_c/2$, in fundamental flux
units, namely,
\beq
{\cal N} = \frac{\langle \Phi(\lambda_c/2)^2 \rangle^{1/2}}{\alpha^{-1/2}} =
(K\alpha \lambda_c)^{1/2}\left( \frac{\pi}{4} {\cal F}(2) \right)^{1/2} \approx
0.46(K\alpha \lambda_c)^{1/2}
\eeq

\section{The confining potential $V(\Phi)$.}
\label{sec:potential}
We wish to choose a potential $V(\Phi)$ which will confine the field by
driving the flux through each plaquette to an integer multiple of
$\Phi_0$.  We treat any multiple the same as any other, so
$V$ should be periodic with period $\Phi_0$.  Similarly, upward and
downward flux are to be treated the same, so $V(\Phi)=V(-\Phi)$.

We choose an inverted parabola centered at $\Phi_0/2$, so that the
flux will be driven to the nearest multiple of $\Phi_0$.  If we continue
this parabola all the way to zero, we find that the flux jumps back and
forth around zero, instead of settling smoothly to zero.  Thus we use a
smooth quadratic form for $V(\Phi)$ near zero.  The smoothing
extends to a critical value $\Phi_*= 1/N^2$.  Any larger value
can result in one unit flux being spread out throughout the lattice
instead of being concentrated on a single plaquette.

\begin{figure}
\centering\leavevmode
\epsfysize=7cm \epsfbox{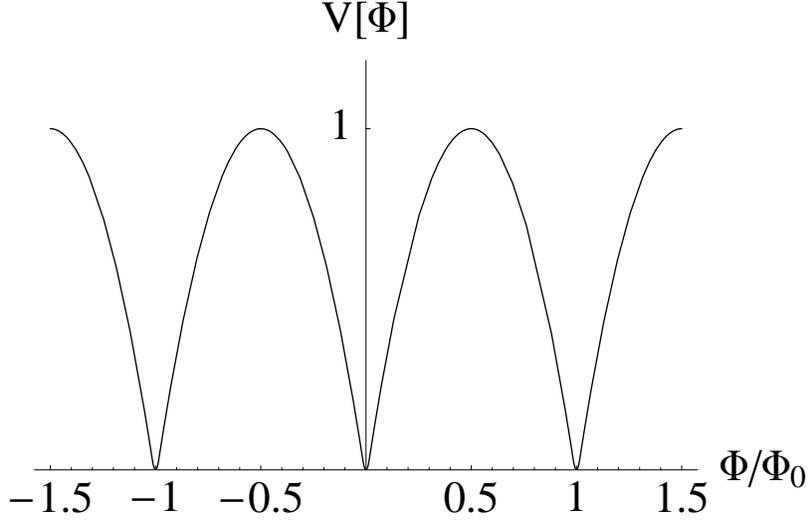}
\caption{Confining potential as a function of the magnetic flux through a single plaquette.}
\label{POTENTIAL}
\end{figure}

Our potential is then given by
\beq
V(\Phi) = \begin{cases}c_1\Phi^2 & 0<\Phi<\Phi_*\\
1-c_2\left(\Phi_0/2-\Phi\right)^2 & \Phi_*<\Phi <\Phi_0/2
\end{cases}
\eeq
and for other values of $\Phi$ by reflection and periodicity, with
\beq
c_1 = \frac{2}{\Phi_*\Phi_0}
\eeq
\beq
c_2 = \frac{2}{\Phi_0(\Phi_0/2-\Phi_*)}\,,
\eeq
which make $V(\Phi)$ and its derivative continuous at $\Phi_*$.  We plot 
in Fig.~\ref{POTENTIAL} the potential $V(\Phi)$ for $N= 64$.

We then choose the largest time step $\Delta t$ which will not
allow the potential to overshoot 0 in a single step.

\end{document}